\documentclass[manuscript,nonacm]{acmart}
\usepackage{multirow}
\usepackage{tabularx}
\usepackage{amsmath}
\usepackage{amsfonts}
\usepackage{algorithm}
\usepackage{algorithmic}
\usepackage{graphicx}
\usepackage{booktabs}
\usepackage{longtable}
\usepackage{array}
\usepackage{enumitem}
\usepackage{float}
\usepackage{siunitx}
\usepackage{tikz}
\usetikzlibrary{arrows.meta,backgrounds,fit,positioning,shapes.geometric,shapes.symbols}
\AtBeginDocument{%
  }

\setcopyright{none}
\renewcommand\footnotetextcopyrightpermission[1]{}

\begin{document}

\title{Defensible Design for OpenClaw: Securing Autonomous Tool-Invoking Agents}

\author{Zongwei Li}

\affiliation{%
  \institution{Hainan University}
  \city{Haikou}
  \country{China}
}
\email{lizw1017@hainanu.edu.cn}

\author{Wenkai Li}
\affiliation{%
  \institution{Hainan University}
  \city{Haikou}
  \country{China}
}
\email{cswkli@hainanu.edu.cn}

\author{Xiaoqi Li}
\authornote{Corresponding author.}
\affiliation{%
  \institution{Hainan University}
  \city{Haikou}
  \country{China}
}
\email{csxqli@ieee.org}
% \author{Zeng Zhang}
% \affiliation{%
%   \institution{Hainan University}
%   \city{Haikou}
%   \country{China}
% }
% \email{zz1up@hainanu.edu.cn}

% \author{Lei Xie}
% \affiliation{%
%   \institution{Hainan University}
%   \city{Haikou}
%   \country{China}
% }
% \email{xielei@hainanu.edu.cn}

\renewcommand{\shortauthors}{Li et al.}

\begin{abstract}
OpenClaw-like agents offer substantial productivity benefits, yet they are insecure by default because they combine untrusted inputs, autonomous action, extensibility, and privileged system access within a single execution loop. We use OpenClaw as an exemplar of a broader class of agents that interact with interfaces, manipulate files, invoke tools, and install extensions in real operating environments. Consequently, their security should be treated as a software engineering problem rather than as a product-specific concern. To address these architectural vulnerabilities, we propose a blueprint for defensible design. We present a risk taxonomy, secure engineering principles, and a practical research agenda to institutionalize safety in agent construction. Our goal is to transition the community focus from isolated vulnerability patching toward systematic defensive engineering and robust deployment practices.
\end{abstract}





\maketitle

\section{Introduction}
OpenClaw illustrates a rapidly emerging class of environment-interactive agents that extend beyond text generation. These systems browse web interfaces, manipulate local files, invoke external tools, and operate within deployment environments that may carry real organizational privileges~\cite{anthropic2024computeruse,openai2025operator,deepmindMariner,chen2026WhenOpenClaw,Weidener2026FromAgentOnlySocial}. For software engineering researchers and practitioners, this shift creates a familiar yet still under-specified systems problem. The central challenge concerns not only model capability, but also the design of architectures, permission boundaries, isolation mechanisms, extension governance, and deployment practices for agents that can translate natural-language goals into consequential actions.

Our starting point is that OpenClaw-like agents are useful, but insecure by default. These agents combine four properties that are individually manageable but jointly destabilizing. They ingest untrusted content from webpages, documents, screenshots, and local artifacts. They autonomously continue tasks rather than merely recommend actions. They are extensible through skills, plugins, or tool integrations. They often execute with access to files, credentials, APIs, and operating system functionality~\cite{anthropic2024computeruse,owaspPromptInjection,openai2025operator}. This combination suggests that prompt injection, harmful misoperation, malicious extensions, and conventional deployment weaknesses should be understood as interacting engineering risks rather than as isolated security anecdotes~\cite{wang2026AssistantDouble,chen2026TrajectoryBasedSafety}.

This framing clarifies why OpenClaw should be treated as an exemplar rather than a singleton case. Prior warnings regarding browser navigation and tool invocation point to a broader threat model where indirect instructions, overbroad permissions, weak isolation, and immature extension governance amplify one another~\cite{owaspPromptInjection,nistAirmf,manik2026OpenClawAgents}. From an engineering perspective, the problem demands defensible design. Developers must build environment interactive agents using architectural safeguards that guarantee operational boundaries remain explicit, governable, and auditable even under malicious instruction.

Rather than claiming completed empirical validation, we synthesize a defensible design framework for studying these agents systematically. We outline the core risk categories that shape the problem layout, distill secure engineering principles for defensive system architecture, and propose a research agenda targeting evaluation, permission management, extension trust, and oversight. Our aim is to turn scattered warnings about agent misuse into a coherent and defensible software engineering program for autonomous systems.

\begin{figure}[t]
\centering
\resizebox{0.7\columnwidth}{!}{%
\begin{tikzpicture}[
  node distance=5mm and 6mm,
  >=Latex,
  font=\scriptsize\sffamily,
  flow/.style={draw, rounded corners=3pt, align=center, minimum width=2.2cm, minimum height=0.78cm, thick},
  io/.style={flow, fill=blue!10, draw=blue!60},
  core/.style={flow, fill=orange!10, draw=orange!60},
  exec/.style={flow, fill=green!10, draw=green!60},
  gov/.style={flow, fill=gray!15, draw=gray!60},
  llm/.style={cloud, cloud puffs=11, cloud puff arc=120, aspect=2.1, draw=purple!60, fill=purple!10, thick, minimum width=2.8cm, minimum height=1.2cm, align=center},
  workspace/.style={cylinder, shape border rotate=90, aspect=0.25, draw=orange!60, fill=orange!10, thick, minimum width=1.8cm, minimum height=1.3cm, align=center},
  arrow/.style={-{Latex[length=2.1mm]}, thick, draw=black!70},
  monitor/.style={-{Latex[length=2.1mm]}, dashed, thick, draw=gray!80},
  bidir/.style={<->, >=Latex, thick, draw=black!70}
]

\node[core] (agent) {OpenClaw Agent Core\\(State \& Reasoning)};
\node[core, below=of agent] (plan) {Plan Selection \&\\Action Decision};
\node[exec, below=of plan] (tools) {Skills, Plugins\\\& Tool Invocation};
\node[exec, below=of tools] (runtime) {Browser, OS, APIs,\\Local Sandbox};

\node[llm, above=of agent, yshift=0.15cm] (model) {Cloud LLM API\\(GPT/Claude/Gemini)};

\node[io, left=of agent, yshift=0.35cm] (user) {User Requests};
\node[io, left=of plan] (inputs) {Webpages, Docs,\\Screenshots, Files};
\node[workspace] (storage) at (inputs |- tools) {Local Workspace\\Memory};

\node[gov, right=of plan] (oversight) {Permissions,\\Oversight, Logging};
\node[gov, right=of runtime] (feedback) {Runtime Feedback\\State Updates};

\draw[bidir] (agent.north) -- node[right, font=\tiny] {Prompt / response} (model.south);

\draw[arrow] (user.east) -- (agent.west);
\draw[arrow] (inputs.east) -- (plan.west);

\draw[bidir] (storage.east) -- node[below, font=\tiny] {Read / write} (tools.west);

\draw[arrow] (agent) -- (plan);
\draw[arrow] (plan) -- (tools);
\draw[arrow] (tools) -- (runtime);

\draw[arrow] (runtime.east) -- (feedback.west);
\draw[arrow] (feedback.north) |- node[above right, font=\tiny] {Update state} (agent.east);

\draw[monitor] (oversight.west) -- (plan.east);
\draw[monitor] (oversight.south west) -- (tools.east);
\draw[monitor] (oversight.south) -- (runtime.north east);

\begin{scope}[on background layer]
  \node[draw=gray!50, fill=gray!3, dashed, rounded corners=8pt, fit=(user) (inputs) (storage) (agent) (plan) (tools) (runtime) (oversight) (feedback), inner sep=10pt, label={[font=\bfseries\scriptsize, text=gray!60, xshift=8mm]above left:Local Environment}] (machine) {};
\end{scope}

\end{tikzpicture}
}

\caption{Architecture and control loop of an OpenClaw-like agent. The agent core coordinates user-facing requests, mixed-trust inputs, local workspace state, cloud LLM interaction, tool invocation, runtime execution, and governance signals within a single operational loop.}
\Description{An architecture diagram for an OpenClaw-like agent. The OpenClaw agent core sits at the center, connected upward to a cloud LLM API, leftward to user chat and gateway requests, mixed-trust inputs, and a local workspace containing SOUL.md and memory, downward to plan selection, tool invocation, and runtime execution in the browser, operating system, APIs, and a local sandbox, and rightward to permissions, oversight, logging, and runtime observations with environment feedback. A dashed background box marks the local environment on the device or VPS.}
\label{fig:openclaw-architecture}
\end{figure}
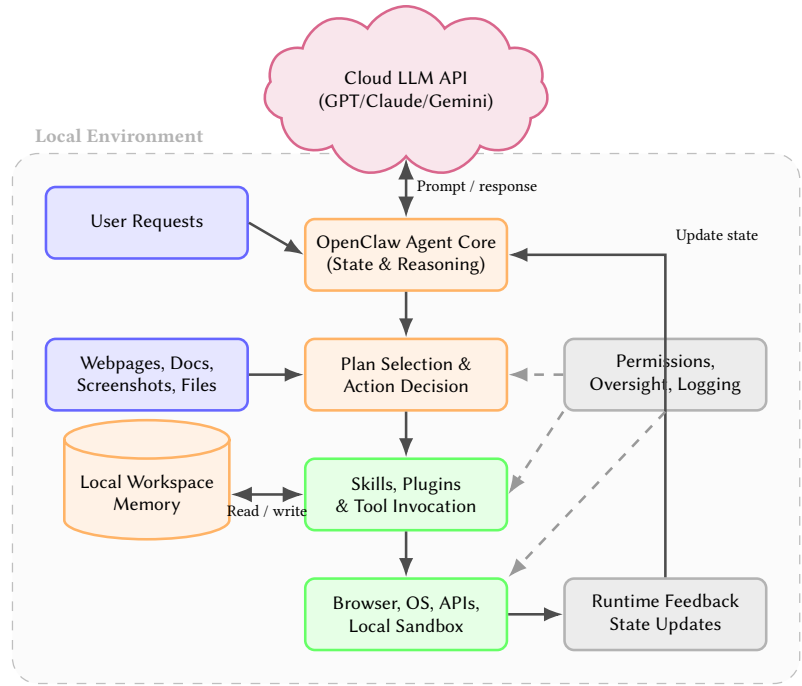

Figure~\ref{fig:openclaw-architecture} summarizes the architecture and control loop of an OpenClaw-like agent. User requests, mixed-trust inputs, local workspace state, cloud LLM interaction, tool invocation, runtime execution, and oversight all participate in the same cycle.

\section{Background}
OpenClaw-like agents are best understood within a broader shift from chat-centered assistants to environment-interactive agents. Rather than mainly generating text, these systems can interpret interfaces, preserve task state across multiple steps, invoke external tools, and act on files, browsers, APIs, and operating-system resources. This shift changes the engineering problem in an important way: capability can no longer be separated from execution context, because the same model output may produce real actions under specific permissions and runtime constraints.

The surrounding ecosystem is similarly diverse and fast-moving. Open-source projects differ in implementation language, deployment target, and trust model, ranging from lightweight single-binary assistants to extensible multi-agent platforms with memory, plugin support, and remote tool connectivity. This diversity matters for secure engineering because risk does not come from the foundation model alone. It also depends on how agent frameworks allocate authority, isolate execution, govern extensions, and expose interfaces and capabilities to users and third-party integrations.

These observations frame the rest of the paper in two ways. First, OpenClaw is better viewed as a representative example of a broader class of agent systems than as an isolated artifact. Second, any useful security analysis must account for the ecosystem in which such systems are built and deployed. We therefore begin with a concise ecosystem snapshot before turning to the risk taxonomy and the secure engineering principles derived from it.

\subsection{OpenClaw-like Ecosystem Snapshot}
Table~\ref{tab:openclaw-like-ecosystem} provides a curated snapshot of OpenClaw-like agent projects spanning multiple implementation languages and deployment styles. The table is organized by implementation language to show ecosystem breadth rather than imply a quality ranking.

OpenClaw should therefore be understood not only as a standalone assistant, but also as one instance in a broader OpenClaw-like ecosystem that includes self-hosted agents, plugin and skill frameworks, multi-agent orchestration platforms, community-maintained extensions, and increasingly varied deployment targets. Within this ecosystem, projects differ not only in implementation language, but also in their assumptions about autonomy, memory, extensibility, interaction channels, and runtime isolation. Some systems emphasize lightweight personal assistance, whereas others focus on multi-agent coordination, local deployment, mobile integration, or resource-constrained execution.

Formal academic literature on this ecosystem remains limited; accordingly, we draw on publicly available repositories, documentation, project artifacts, and a small number of early analyses to support this ecosystem view. Existing analyses describe OpenClaw and related systems as part of a broader sociotechnical environment in which agent platforms, shared infrastructures, and organizational forms evolve together~\cite{Weidener2026FromAgentOnlySocial}. Work on Moltbook further suggests that OpenClaw-style agents can participate in agent-only communities that exchange tutorials, operational knowledge, and behavioral norms~\cite{chen2026WhenOpenClaw,manik2026OpenClawAgents}. Taken together, these sources suggest that the OpenClaw-like landscape is not merely a collection of repositories. It is also an ecosystem of reusable prompts, skills, tools, extensions, interaction patterns, and social feedback mechanisms that circulate across implementations.

From this perspective, the significance of Table~\ref{tab:openclaw-like-ecosystem} lies not only in the number of available projects, but also in the fact that they occupy a shared design space. Across languages and platforms, recurring dimensions include plugin and skill mechanisms, memory and tool integration, local versus remote execution, interaction surfaces such as terminals, chat interfaces, browsers, and messaging platforms, and the extent to which agents are embedded in collaborative or community settings. This diversity helps explain why OpenClaw is more usefully treated as a representative instance of a broader class of systems. It also provides the background for the remainder of the paper, which abstracts from any single implementation to analyze common design tensions and their downstream security implications.

\begingroup
% \footnotesize
\setlength{\tabcolsep}{4pt}
\renewcommand{\arraystretch}{1.12}
\begin{table*}[ht]
\caption{Curated OpenClaw-like agent ecosystem with repository references.}
\label{tab:openclaw-like-ecosystem}
\centering
\begin{tabularx}{\textwidth}{>{\raggedright\arraybackslash}p{0.17\textwidth} >{\raggedright\arraybackslash}p{0.14\textwidth} X}
\toprule
Project & Language & Positioning \\
\midrule
OpenClaw~\cite{repoOpenClaw} & TypeScript & Full-featured assistant with multi-agent routing, voice, and broad channel support. \\
TinyClaw~\cite{repoTinyClaw} & Shell/TypeScript & Workspace-isolated multi-agent runner for chained team workflows. \\
NanoClaw~\cite{repoNanoClaw} & TypeScript & Container-sandboxed assistant built for simplicity and skill-based extension. \\
BabyClaw~\cite{repoBabyClaw} & JavaScript & Telegram-first single-file assistant with voice and scheduled automation. \\
droidclaw~\cite{repoDroidClaw} & TypeScript & Android-oriented OpenClaw variant for lightweight mobile automation. \\
Flowly AI~\cite{repoFlowlyAI} & TypeScript & Flow-centric framework for tool-driven agent workflows. \\
troublemaker~\cite{repoTroublemaker} & TypeScript & Minimal cross-platform runtime for personal agents. \\
LettaBot~\cite{repoLettaBot} & TypeScript & Persistent-memory assistant across major chat platforms. \\
SupaClaw~\cite{repoSupaClaw} & TypeScript & Supabase-native assistant stack for self-hosted deployment. \\
ZeroClaw~\cite{repoZeroClaw} & Rust & Trait-based low-overhead infrastructure with a swappable core. \\
Moltis~\cite{repoMoltis} & Rust & Single-binary gateway with memory, sandboxing, and multi-channel access. \\
IronClaw~\cite{repoIronClaw} & Rust & Privacy-first assistant with encrypted local state and layered security. \\
ZeptoClaw~\cite{repoZeptoClaw} & Rust & Tiny hardened runtime emphasizing isolation and secret scanning. \\
shrew~\cite{repoShrew} & Rust & Compact high-speed runtime for lightweight autonomy. \\
moxxy~\cite{repoMoxxy} & Rust & Self-hosted Rust framework for multi-agent systems. \\
Microclaw~\cite{repoMicroclaw} & Rust & Chat-surface assistant runtime inspired by NanoClaw. \\
OpenFang~\cite{repoOpenFang} & Rust & Modular Rust agent OS for larger-scale orchestration. \\
nanobot~\cite{repoNanobot} & Python & Lightweight research assistant with one-click deployment and MCP. \\
HermitClaw~\cite{repoHermitClaw} & Python & Folder-resident autonomous researcher producing reports, notes, and scripts. \\
AstrBot~\cite{repoAstrBot} & Python & Plugin-driven IM agent platform with broad messaging support. \\
safeclaw~\cite{repoSafeClaw} & Python & Safety-oriented text-and-voice assistant with reduced LLM reliance. \\
Clawlet~\cite{repoClawlet} & Python & Identity-aware lightweight framework for rapid setup. \\
AngelClaw~\cite{repoAngelClaw} & Python & Lean research-oriented OpenClaw variant for experimental ideas. \\
pickle-bot~\cite{repoPickleBot} & Python & Lightweight self-hosted personal assistant. \\
PicoClaw~\cite{repoPicoClaw} & Go & Single-binary assistant for constrained hardware and older phones. \\
picobot~\cite{repoPicobot} & Go & Minimal self-hosted Go bot in a single binary. \\
MimiClaw~\cite{repoMimiClaw} & C & ESP32-S3 local-first assistant for always-on low-power use. \\
zclaw~\cite{repoZclaw} & C & Minimal ESP32 assistant runtime for ultra-small deployments. \\
subzeroclaw~\cite{repoSubZeroClaw} & C & Skill-driven edge daemon for embedded workflows. \\
NullClaw~\cite{repoNullClaw} & Zig & Tiny portable infrastructure for resource-constrained assistants. \\
Autobot~\cite{repoAutobot} & Crystal & Kernel-sandboxed assistant with multi-provider, voice, vision, and MCP. \\
\bottomrule
\end{tabularx}
\end{table*}
\endgroup

\section{Risk Taxonomy}
This section organizes the threat model into four risk classes and situates them along the execution path of an OpenClaw-like agent, from mixed-trust inputs to deployment weaknesses. To complement the four analytic classes below, Table~\ref{tab:pipeline-risk-surface} reframes the attack surface by operational stage. Rather than listing isolated incidents, it highlights a compact set of trust transitions where security failures typically accumulate. This operational view does not replace the four risk classes below. Instead, it makes their placement explicit across the pipeline. The second column identifies the dominant class for each stage, while recognizing that some incidents may overlap multiple classes in practice.

\begin{table}[t]
\caption{Pipeline-oriented reframing of the risk surface for OpenClaw-like agents.}
\label{tab:pipeline-risk-surface}
% \footnotesize
\setlength{\tabcolsep}{4pt}
\renewcommand{\arraystretch}{1.15}
\begin{tabularx}{\columnwidth}{>{\raggedright\arraybackslash}p{0.21\columnwidth} >{\raggedright\arraybackslash}p{0.20\columnwidth} >{\raggedright\arraybackslash}p{0.21\columnwidth} X}
\toprule
Stage & Primary taxonomy class & Reframed risk surface & Illustrative failures \\
\midrule
Message ingestion and adaptation & Prompt Injection & Instruction-channel compromise & Hidden directives in text, images, or metadata survive preprocessing and are treated as legitimate task instructions. \\
Gateway authentication and routing & Deployment Vulnerabilities & Control-plane trust failure & Weak authentication, exposed endpoints, or incorrect session binding let unauthorized requests inherit valid execution context. \\
Context assembly and memory use & Prompt Injection & Persistent context corruption & Poisoned history, retrieved notes, or cached state bias later planning and keep unsafe instructions alive across turns. \\
Runtime planning and task continuation & Harmful Misoperation & Authority drift & The agent overgeneralizes user intent, expands task scope, or continues acting under uncertainty until benign ambiguity becomes harmful misoperation. \\
Tool and command execution & Harmful Misoperation & Execution-plane abuse & Over-privileged tools, unsafe command construction, or unchecked external actions turn model output into direct system compromise. \\
Skill, plugin, and dependency loading & Extension Supply-Chain Risk & Extensibility trust failure & Malicious skills, compromised plugins, or unverified remote dependencies expand the trusted computing base and create persistence paths. \\
Response generation and delivery & Deployment Vulnerabilities & Sensitive egress and propagation & Secrets, local state, or unsafe content are returned outward and then relayed into downstream chats, platforms, or workflows. \\
Logging and persistent storage & Deployment Vulnerabilities & Forensic and state contamination & Attacker-controlled content distorts logs, weakens auditability, and is retained in long-term stores for reuse in later sessions. \\
\bottomrule
\end{tabularx}
\end{table}

\subsection{Prompt Injection}
Prompt injection begins at the instruction layer, where mixed-trust external content alters the agent's effective control flow. Hidden text in a webpage, document, email, screenshot, or local note can operate as an indirect command that competes with the user's stated objective and steers the agent toward unauthorized actions. The exposure is structural because OpenClaw-like agents continuously ingest heterogeneous content while retaining the ability to browse, read files, invoke tools, and continue tasks with limited interruption. In a text-only system, malicious content may corrupt an answer. In an environment-interactive agent, the same content can redirect an autonomous and privileged execution loop toward concrete actions such as secret retrieval, unsafe navigation, or unauthorized data transfer.

In practice, this failure mode includes following hidden instructions embedded in webpages, treating attacker-controlled text as higher-priority guidance than the user's goal, disclosing environment-derived secrets, or executing unintended file and network operations under the guise of task completion. The core issue is not hallucination, but indirect instruction hijacking through untrusted inputs that the agent is expected to parse during normal work. The resulting damage varies by setting. For personal users, this risk can expose private files, credentials, chats, and payment-linked information. In enterprise settings, it can leak code, documents, API tokens, or internal workflow state. In critical domains, the same mechanism can misroute sensitive actions or degrade operational integrity when agents are attached to high-consequence systems.

\subsection{Harmful Misoperation}
Unlike prompt injection, harmful misoperation does not require an attacker. It arises when the agent translates an underspecified, ambiguous, or only partially observed goal into an irreversible action sequence that the user did not actually intend. This risk is pronounced because OpenClaw-like agents operate across interfaces, files, and tools in contexts where intent must be inferred from incomplete information and where the cost of a wrong action can be immediate. A text-only system can misunderstand a request and return a poor answer. An environment-interactive agent can misunderstand the same request and then delete data, modify records, send messages, or trigger workflows using the user's effective privileges.

Common manifestations include deleting or overwriting the wrong file set, sending information to unintended recipients, executing the wrong workflow branch after misreading interface state, or persisting a mistaken action because the environment offers weak opportunities for recovery. These errors are especially likely when instructions are vague, interface state is partially hidden, or the task spans multiple dependent steps~\cite{chen2026TrajectoryBasedSafety}. The consequences can be substantial even without adversarial input. In enterprise settings, harmful misoperation can corrupt documents, disrupt operational workflows, leak sensitive records, or create audit and compliance exposure. For personal users, it can destroy messages, photos, notes, or financial data. In critical domains, the same pattern can translate into service disruption, incorrect configuration changes, or other high-consequence operational mistakes.

\subsection{Extension Supply-Chain Risk}
Extensibility introduces a distinct supply-chain risk when a skill, plugin, tool wrapper, packaged workflow, or update channel becomes the path through which adversarial logic enters the agent stack. The problem is not merely that an extension can be buggy, but that extensibility moves new code, prompts, permissions, and execution assumptions into the trusted computing base. The exposure grows because OpenClaw-like agents are designed to derive utility from integration. They call tools, install extensions, and compose workflows that inherit the privileges of the surrounding agent runtime. A text-only system may echo or recommend malicious extension content. An environment-interactive agent can load that content into an autonomous execution loop that can browse, read, write, and invoke services with real authority. In this setting, autonomy and privilege amplification make the risk materially more severe.

Concrete failures include installing or invoking a malicious plugin that exfiltrates local data, routing tasks through a compromised tool wrapper that silently broadens access, accepting an update that changes execution behavior in attacker-favorable ways, or importing a workflow component whose hidden instructions subvert the user task~\cite{Clawdrain,manik2026OpenClawAgents}. Each added integration enlarges the trust boundary and makes it harder to reason about what the agent is actually authorized to do. Once compromised, the impact can spread quickly across environments. For personal users, supply-chain compromise can expose files, browser state, and linked accounts through seemingly helpful extensions. In enterprise settings, it can leak source code, documents, credentials, or workflow metadata while providing an attacker with a durable foothold inside routine automation. In critical domains, compromised extensions can contaminate operator tooling or decision-support pathways, turning one bad dependency into system-wide operational exposure.

\subsection{Deployment Vulnerabilities}
Beyond model behavior, the agent runtime inherits the conventional security weaknesses of deployed software systems. These include exposed services, weak authentication, insecure defaults, secret leakage, weak isolation, dependency flaws, and other vulnerabilities that allow attackers to reach or abuse the deployed system. Although the model is the most visible component, the operational attack surface also includes web front ends, APIs, containers, host permissions, credential stores, and background services. This exposure persists because OpenClaw-like agents are deployed as long-lived, connected systems that bridge model reasoning with browsers, tools, files, and external services. A text-only chatbot with limited authority may be constrained to producing harmful text. An environment-interactive agent running with network reachability, stored secrets, and ambient operating privileges turns ordinary software weaknesses into direct paths for commandeering real-world actions at machine speed.

Operationally, the failures are familiar but more consequential. Representative cases include exposing an agent endpoint without strong authentication, shipping permissive default settings that overgrant file or tool access, leaking API tokens or session material through logs and configuration artifacts, or relying on weak sandboxing that allows one compromised component to pivot into the broader runtime. These are not generic model-safety failures, but deployment and runtime security problems made more consequential by an agent that can immediately act on whatever access it receives. At the system level, the resulting damage can be extensive. In enterprise settings, this class can expose internal services, documents, credentials, and automation pathways while making incident containment harder once the agent is embedded in daily operations. For personal users, insecure deployment can reveal private data or give attackers remote control over local workflows. In critical domains, weak isolation or exposed administrative surfaces can convert a routine software flaw into service disruption, safety-impacting misconfiguration, or broader infrastructure compromise.

Taken together, the four categories show why environment-interactive agent risk cannot be reduced to any single failure narrative. Personal users may experience privacy loss, account compromise, or destructive automation. Enterprises face the combined possibility of data leakage, workflow corruption, persistent footholds, and audit failures. Critical-domain operators inherit the additional danger that compromised autonomy can propagate into high-consequence environments. This cross-setting pattern makes enterprise deployment the natural anchor while still requiring consequence reasoning that extends to consumer and critical infrastructures.

The next step, therefore, is not to extend the taxonomy further, but to derive secure engineering principles that map directly to these four risk classes.

\begin{figure}[t]
\centering
\resizebox{0.8\columnwidth}{!}{%
\begin{tikzpicture}[
  node distance=5mm and 9mm,
  >=Latex,
  font=\scriptsize\sffamily,
  risk/.style={draw, rounded corners=3pt, align=center, minimum width=4.2cm, minimum height=0.8cm, thick, fill=red!8, draw=red!55},
  ctrl/.style={draw, rounded corners=3pt, align=center, minimum width=3.9cm, minimum height=0.8cm, thick, fill=blue!8, draw=blue!55},
  head/.style={font=\bfseries\scriptsize, align=center},
  pmap/.style={-{Latex[length=2.0mm]}, very thick, draw=black!75},
  smap/.style={-{Latex[length=1.9mm]}, thick, dashed, draw=gray!70}
]

\node[head] (lh) at (0,0) {Risk Taxonomy};

\node[risk, below=of lh, yshift=-0.5mm] (r1) {Prompt Injection};
\node[risk, below=of r1] (r2) {Harmful Misoperation};
\node[risk, below=of r2] (r3) {Extension Supply-Chain Risk};
\node[risk, below=of r3] (r4) {Deployment Vulnerabilities};

\node[ctrl, right=of r1, xshift=12mm] (c1) {Least Privilege};
\node[ctrl, below=of c1] (c2) {Runtime Isolation};
\node[ctrl, below=of c2] (c3) {Extension Governance};
\node[ctrl, below=of c3] (c4) {Auditability};
\node[head] (rh) at (c1 |- lh) {Secure Engineering Principles};

% === 主要缓解路径 (Primary Mitigation Paths) ===
\draw[pmap] (r1.east) -- (c1.west);
\draw[pmap] (r2.east) -- (c1.west);
\draw[pmap] (r3.east) -- (c3.west);
\draw[pmap] (r4.east) -- (c2.west);

% === 次要/辅助缓解路径 (Secondary Mitigation Paths) ===
\draw[smap] (r1.east) -- (c2.west);
\draw[smap] (r1.east) -- (c4.west); % 新增：提示词注入 -> 可审计性
\draw[smap] (r2.east) -- (c2.west); % 新增：有害误操作 -> 运行时隔离
\draw[smap] (r2.east) -- (c4.west);
\draw[smap] (r3.east) -- (c2.west);
\draw[smap] (r3.east) -- (c4.west);
\draw[smap] (r4.east) -- (c4.west); % 修改：部署漏洞 -> 可审计性 (改为虚线)

\end{tikzpicture}
}
\caption{Risk-to-control mapping for OpenClaw-like agents. The taxonomy on the left is connected to the engineering principles on the right through primary and secondary mitigation paths.}
\Description{A two-column mapping diagram. The left column lists four risk classes: Prompt Injection, Harmful Misoperation, Extension Supply-Chain Risk, and Deployment Vulnerabilities. The right column lists four engineering principles: Least Privilege, Runtime Isolation, Extension Governance, and Auditability. Solid arrows indicate primary mitigation paths. Dashed arrows indicate secondary influences.}
\label{fig:risk-control-mapping}
\end{figure}
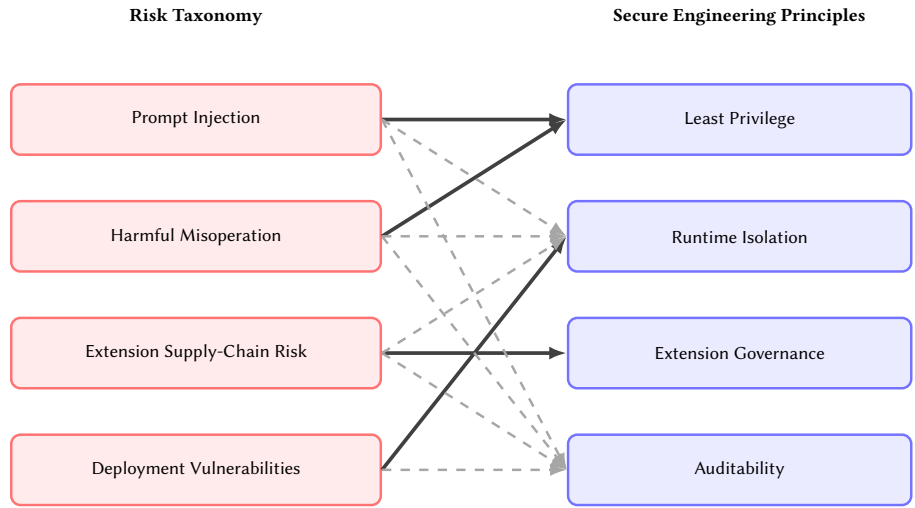

Figure~\ref{fig:risk-control-mapping} summarizes this mapping. The left side captures the four dominant risk classes, while the right side highlights the engineering principles that primarily constrain each class. Solid arrows denote primary preventative paths that reduce exploitability or blast radius, while dashed arrows denote secondary or detective influence that improves monitoring, attribution, and recovery.

\section{Secure Engineering Principles}
The taxonomy above matters only if it leads to actionable design guidance. For OpenClaw-like agents, secure engineering begins with bounded authority and controlled runtime exposure.

\subsection{Least Privilege}
Least privilege and capability bounding are foundational controls for environment-interactive agents because these systems routinely browse, read files, invoke tools, and continue tasks without human re-authorization at each step. The principle is straightforward. An agent should receive only the minimum authority, resources, and action scope required for the current task, and those permissions should remain bounded even when the agent encounters unexpected inputs or only partially understands user intent.

This principle directly addresses prompt injection and indirect instruction hijacking by preventing attacker-controlled content from escalating into broad file, network, or account access. It also addresses harmful misoperation and over-autonomy by limiting the damage that a mistaken plan can cause when the agent takes the wrong branch. This constraint matters operationally because environment-interactive agents do not fail like static software components. They fail while navigating interfaces, selecting tools, and chaining actions across environments. If capabilities are ambient and overbroad, one incorrect inference or poisoned instruction can quickly propagate into destructive or irreversible outcomes. Treating least privilege as a baseline design principle therefore preserves utility while preventing task continuation from becoming unchecked delegated power.

\subsection{Runtime Isolation}
Runtime isolation and secret hygiene are deployment necessities for environment-interactive agents because these systems execute in environments where browsing, tool use, file access, and stored credentials can all become part of the same runtime path. The principle is that the agent should operate within isolation boundaries that constrain what each session, tool, and extension can reach. Secrets should be scoped, exposed only when necessary, and withheld from the agent's ambient context by default.

This principle directly addresses traditional software vulnerabilities and insecure deployment configuration by limiting how far ordinary runtime weaknesses can spread. It also addresses prompt injection, indirect instruction hijacking, and malicious skills, plugins, or supply-chain compromise, because an injected instruction or compromised extension becomes substantially more dangerous when it can freely access tokens, local state, or adjacent services. This matters operationally because environment-interactive agents are rarely single-process abstractions. They are deployed systems that persist across sessions, connect to external services, and accumulate credentials, logs, and artifacts over time. Without meaningful isolation and disciplined secret handling, one compromised browser session, one over-privileged tool invocation, or one insecure extension can turn a local failure into a broader organizational breach. Treating isolation and secret hygiene as deployment fundamentals therefore prevents autonomous capability from inheriting the full blast radius of the runtime that hosts it.

\subsection{Extension Governance}
Extension trust, provenance, and permission governance should be treated as first-class engineering principles because environment-interactive agents derive much of their usefulness from skills, plugins, tool wrappers, packaged workflows, and update channels that enlarge the trusted computing base. The core claim is that extensibility is not a convenience layer outside the security model. Every extension path imports new prompts, code, permissions, and behavioral assumptions that may inherit sensitive authority from the surrounding runtime.

This principle directly addresses malicious skills, plugins, and supply-chain compromise by making provenance, trust, and bounded delegation explicit design concerns. It also addresses damage amplification when overbroad permissions allow a compromised extension to reach files, services, or workflows far beyond its legitimate task. This matters operationally because environment-interactive agents do not merely recommend third-party actions. They can invoke extensions inside an autonomous loop that already has the ability to browse, read, write, and act. If provenance is weak or permission governance is implicit, one malicious or silently modified dependency can become a durable path for exfiltration, workflow manipulation, or privilege abuse. Treating extension governance as a baseline principle therefore keeps extensibility compatible with secure operation rather than allowing ecosystem growth to become an uncontrolled expansion of agent authority.

\subsection{Auditability}
Logging, auditability, and forensic visibility are necessary engineering principles for autonomous environment-interactive agents because these systems take consequential actions across mixed-trust inputs, delegated tools, extensions, and mutable runtime environments. The principle is that an agent should leave observable and attributable traces that make it possible to determine what it saw, which decision path it followed, what authority it used, and which external component influenced the outcome.

This principle addresses the earlier risk classes collectively rather than individually. Prompt injection and harmful misoperation are difficult to distinguish after the fact without decision traces. Extension abuse and supply-chain compromise are difficult to attribute without visibility into invoked components and provenance. Deployment or runtime compromise is harder to contain when the system provides only ordinary debugging logs rather than security-relevant evidence. This matters operationally because failures in environment-interactive agents often unfold as multi-step episodes rather than isolated bugs. When an agent reads mixed-trust content, calls tools, traverses interfaces, and acts with delegated authority, post-incident recovery depends on reconstructing how control flowed and where the boundary failed. In this sense, auditability is primarily a detective and accountability control that complements, rather than replaces, preventative controls such as least privilege and runtime isolation. Treating auditability as a baseline principle therefore supports accountability, incident response, and defensible deployment rather than leaving operators blind once autonomous execution goes wrong.

\section{Research Agenda}
\begin{figure}[t]
\centering
\resizebox{\textwidth}{!}{%
\begin{tikzpicture}[
  >=Latex,
  font=\scriptsize\sffamily,
  foundation/.style={draw, rounded corners=3pt, thick, align=center, minimum width=3.55cm, minimum height=1.65cm, inner sep=5pt, fill=gray!12, draw=gray!60},
  translate/.style={draw, rounded corners=3pt, thick, text width=2.65cm, minimum height=1.15cm, align=center, inner sep=5pt, fill=orange!10, draw=orange!60},
  agenda/.style={draw, rounded corners=3pt, thick, align=center, minimum width=2.95cm, minimum height=1.65cm, inner sep=5pt, fill=blue!8, draw=blue!60},
  goal/.style={draw, rounded corners=3pt, thick, align=center, text width=2.75cm, minimum height=2.25cm, inner sep=6pt, fill=green!10, draw=green!50},
  group/.style={draw, rounded corners=6pt, dashed, thick, inner sep=0pt},
  connector/.style={-{Latex[length=2.2mm, width=1.3mm]}, thick, draw=black!70, shorten <=3pt, shorten >=3pt},
  head/.style={font=\bfseries\scriptsize, text=black!75}
]

\node[group, draw=gray!50, fill=gray!5, minimum width=3.9cm, minimum height=4cm, label={[head, yshift=2pt]above:Security Framing}] (framing) at (-5.35,0) {};
\node[translate] (translation) at (-1.2,0) {Translate framing into\\actionable engineering\\mechanisms};
\node[group, draw=blue!45, fill=blue!5, minimum width=6.4cm, minimum height=4cm, label={[head, yshift=2pt]above:Research Agenda}] (agendablock) at (4.1,0) {};
\node[goal] (goal) at (9.6,0) {\textbf{Deployment Goal}\\Defensible environment-interactive\\agents should be testable, bounded,\\governable, and auditable};

\node[foundation] (risk) at ([yshift=0.95cm]framing.center) {\parbox[c][1.32cm][c]{3.2cm}{\centering Risk Taxonomy\\Prompt Injection, Misoperation,\\Supply-Chain Risk,\\Deployment Vulnerabilities}};
\node[foundation] (principles) at ([yshift=-0.95cm]framing.center) {\parbox[c][1.32cm][c]{3.2cm}{\centering Secure Engineering Principles\\Least Privilege, Isolation,\\Governance, Auditability}};

\node[agenda] (eval) at ([xshift=-1.55cm,yshift=0.95cm]agendablock.center) {\parbox[c][1.32cm][c]{2.6cm}{\centering \textbf{Evaluation}\\\textbf{Infrastructure}\\Benchmarks, regression testing,\\and metrics for secure agent behavior}};
\node[agenda] (perm) at ([xshift=1.55cm,yshift=0.95cm]agendablock.center) {\parbox[c][1.32cm][c]{2.6cm}{\centering \textbf{Permission}\\\textbf{Architecture}\\Map language-level goals to bounded\\actions and delegated authority}};
\node[agenda] (gov) at ([xshift=-1.55cm,yshift=-0.95cm]agendablock.center) {\parbox[c][1.32cm][c]{2.6cm}{\centering \textbf{Extension}\\\textbf{Governance}\\Provenance, manifests, attestation,\\and revocation for skills and plugins}};
\node[agenda] (oversight) at ([xshift=1.55cm,yshift=-0.95cm]agendablock.center) {\parbox[c][1.32cm][c]{2.6cm}{\centering \textbf{Oversight \&}\\\textbf{Telemetry}\\Risk-adaptive approvals\\with attributable execution traces}};

\draw[connector] (framing.east) -- (translation.west);
\draw[connector] (translation.east) -- (agendablock.west);
\draw[connector] (agendablock.east) -- (goal.west);

\end{tikzpicture}
}
\caption{Overview of the research agenda for OpenClaw-like agents. The earlier security framing is translated into four engineering workstreams: evaluation infrastructure, permission architecture, extension governance, and adaptive oversight with attributable telemetry. Together, these workstreams aim to make agent behavior testable, bounded, governable, and auditable in deployment.}
\Description{A horizontal overview diagram for the research agenda of OpenClaw-like agents. A left panel labeled Security Framing contains two boxes for risk taxonomy and secure engineering principles. An arrow leads to a central translation box. Another arrow leads to a right panel labeled Research Agenda that contains four modules: evaluation infrastructure, permission architecture, extension governance, and oversight with telemetry. A final arrow points to a deployment goal box stating that environment-interactive agents should be testable, bounded, governable, and auditable.}
\label{fig:research-agenda-framework}
\end{figure}
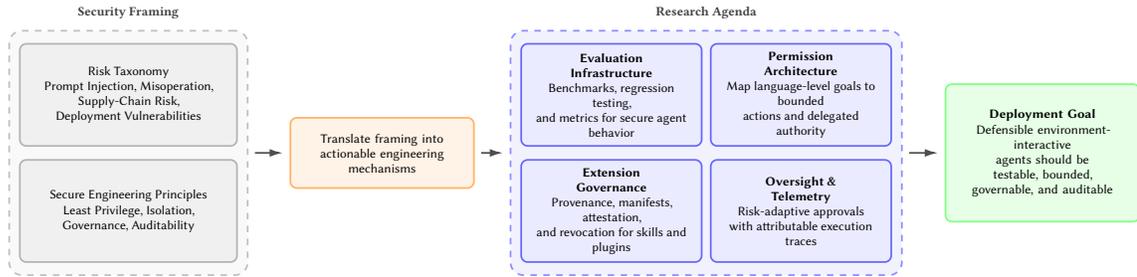

The principles above matter only if they can be translated into a concrete software-engineering program. For OpenClaw-like agents, the next step is not a longer catalog of warnings, but a focused research agenda that connects evaluation infrastructure, control architecture, ecosystem governance, and post-incident visibility. Such an agenda can turn broad security concerns into tractable engineering objectives and, in turn, make autonomous computer use more defensible in practice. Figure~\ref{fig:research-agenda-framework} summarizes this translation from the earlier security framing to four concrete engineering workstreams and a shared deployment goal.

The first research direction concerns benchmark-driven evaluation infrastructure. Secure environment-interactive agents need reproducible benchmark suites that support development, regression testing, and cross-system comparison under realistic operating conditions~\cite{wang2026AssistantDouble,chen2026TrajectoryBasedSafety}. The central open problem is how to evaluate robustness when mixed-trust content, ambiguous tasks, delegated tools, and real permission boundaries interact within the same execution loop. The required artifact is a benchmark infrastructure that measures adversarial robustness, harmful-action prevention, permission-boundary compliance, oversight handoff behavior, and extension or tool risk within representative workflows. The key evaluation criterion is whether such an infrastructure supports repeatable comparison, regression detection, and principled validation of secure-engineering claims, rather than merely collecting isolated attack anecdotes.

The second research direction concerns permission architecture for natural-language-driven action. The open problem is how to translate ambiguous user intent into bounded actions, capabilities, or policy-mediated authority without reverting to ambient privilege. The required artifact is not another generic confirmation prompt. It is a permission architecture that mediates what an agent may do before execution begins and as task context evolves. The key evaluation criterion is whether such a design preserves task utility while making least privilege and capability bounding operational at scale, especially when the agent must move from high-level language instructions to concrete actions over files, tools, services, and interfaces.

The third research direction concerns governance for skills, plugins, and extensions as a platform-design problem. The open problem is expansion of the trusted computing base. Every added skill, plugin, tool wrapper, or packaged workflow imports new code, prompts, permissions, and update channels into the agent stack. The required artifact is a governance layer that verifies publisher identity, declared permissions, provenance, attestation status, and revocation state before delegated authority is inherited. Signed extension manifests represent one plausible design approach. The key evaluation criterion is whether this governance reduces unsafe-extension acceptance and enforces declared boundaries without rendering legitimate extensibility impractical.

The fourth research direction concerns adaptive human oversight together with attributable telemetry. These concerns should be studied together because effective intervention depends on knowing why the agent acted, which authority it used, and which external component shaped the outcome. The open problem is how to trigger intervention for irreversible, high-sensitivity, uncertain, or policy-conflicted actions without overwhelming operators, even when those actions technically remain within an existing permission boundary. The required artifacts are a risk-adaptive approval mechanism and attributable execution tracing that records input provenance, invoked tools and extensions, policy decisions, approvals, authority used, and resulting actions. The key evaluation criterion is whether this combined design reduces avoidable interruptions, intercepts high-risk actions, and enables investigators to distinguish prompt injection, harmful misoperation, and extension abuse with sufficient fidelity for incident response and organizational learning.

Taken together, these research directions translate the earlier principles into a coherent software-engineering agenda. This agenda calls for validation infrastructures for secure behavior, bounded control architectures for action, governance mechanisms for extensible agent ecosystems, and evidentiary foundations for intervention, accountability, and post-incident learning.

\section{Conclusion}
OpenClaw-like agents are highly capable yet insecure by default because they combine mixed-trust inputs, autonomous action, extensibility, and privileged system access within a single operational loop. We argue that securing these systems requires defensible design rather than mere improvements in model reasoning. The proposed threat taxonomy, engineering principles, and research directions form a connected account of how risk enters the agent workflow and how defensible architectures can constrain it. Safer deployment strictly depends on bounded authority, runtime isolation, extension governance, and auditability. Taken together, these elements provide a defensible design blueprint for making autonomous computer use robust and governable in real operating environments.

\bibliographystyle{unsrt} 
\bibliography{references}

@misc{anthropic2024computeruse,
  title = {Introducing computer use, a new Claude 3.5 Sonnet, and Claude 3.5 Haiku},
  author = {{Anthropic}},
  year = {2024},
  howpublished = {\url{https://www.anthropic.com/news/3-5-models-and-computer-use}}
}

@misc{openai2025operator,
  title = {Introducing Operator},
  author = {{OpenAI}},
  year = {2025},
  howpublished = {\url{https://openai.com/index/introducing-operator/}}
}

@misc{owaspPromptInjection,
  title = {LLM Prompt Injection Prevention Cheat Sheet},
  author = {{OWASP}},
  year = {2025},
  howpublished = {\url{https://cheatsheetseries.owasp.org/cheatsheets/LLM_Prompt_Injection_Prevention_Cheat_Sheet.html}}
}

@misc{nistAirmf,
  title = {Artificial Intelligence Risk Management Framework (AI RMF)},
  author = {{National Institute of Standards and Technology}},
  year = {2023},
  howpublished = {\url{https://www.nist.gov/itl/ai-risk-management-framework}}
}

@misc{deepmindMariner,
  title = {Project Mariner},
  author = {{Google DeepMind}},
  year = {2024},
  howpublished = {\url{https://deepmind.google/technologies/project-mariner/}}
}

@misc{Clawdrain,
  title = {Clawdrain: {{Exploiting Tool-Calling Chains}} for {{Stealthy Token Exhaustion}} in {{OpenClaw Agents}}},
  author = {Dong, Ben and Feng, Hui and Wang, Qian},
  year = 2026,
  number = {arXiv:2603.00902},
  eprint = {2603.00902},
  primaryclass = {cs},
  archiveprefix = {arXiv}
}

@misc{Weidener2026FromAgentOnlySocial,
  title = {From {{Agent-Only Social Networks}} to {{Autonomous Scientific Research}}: {{Lessons}} from {{OpenClaw}} and {{Moltbook}}, and the {{Architecture}} of {{ClawdLab}} and {{Beach}}.{{Science}}},
  author = {Weidener, Lukas and Brki{\'c}, Marko and Lee, Phillip and Karlsson, Martin and Noessler, Kevin and Kohlhaas, Paul},
  year = 2026,
  number = {arXiv:2602.19810},
  eprint = {2602.19810},
  primaryclass = {cs},
  archiveprefix = {arXiv}
}

@misc{chen2026TrajectoryBasedSafety,
  title = {A {{Trajectory-Based Safety Audit}} of {{Clawdbot}} ({{OpenClaw}})},
  author = {Chen, Tianyu and Liu, Dongrui and Hu, Xia and Yu, Jingyi and Wang, Wenjie},
  year = 2026,
  number = {arXiv:2602.14364},
  eprint = {2602.14364},
  primaryclass = {cs},
  archiveprefix = {arXiv}
}

@misc{chen2026WhenOpenClaw,
  title = {When {{OpenClaw AI Agents Teach Each Other}}: {{Peer Learning Patterns}} in the {{Moltbook Community}}},
  author = {Chen, Eason and Guan, Ce and Elshafiey, Ahmed and Zhao, Zhonghao and Zekeri, Joshua and Shaibu, Afeez Edeifo and Prince, Emmanuel Osadebe},
  year = 2026,
  number = {arXiv:2602.14477},
  eprint = {2602.14477},
  primaryclass = {cs},
  archiveprefix = {arXiv}
}

@misc{manik2026OpenClawAgents,
  title = {{{OpenClaw Agents}} on {{Moltbook}}: {{Risky Instruction Sharing}} and {{Norm Enforcement}} in an {{Agent-Only Social Network}}},
  author = {Manik, Md Motaleb Hossen and Wang, Ge},
  year = 2026,
  number = {arXiv:2602.02625},
  eprint = {2602.02625},
  primaryclass = {cs},
  archiveprefix = {arXiv}
}

@misc{wang2026AssistantDouble,
  title = {From {{Assistant}} to {{Double Agent}}: {{Formalizing}} and {{Benchmarking Attacks}} on {{OpenClaw}} for {{Personalized Local AI Agent}}},
  author = {Wang, Yuhang and Xu, Feiming and Lin, Zheng and He, Guangyu and Huang, Yuzhe and Gao, Haichang and Niu, Zhenxing and Lian, Shiguo and Liu, Zhaoxiang},
  year = 2026,
  number = {arXiv:2602.08412},
  eprint = {2602.08412},
  primaryclass = {cs},
  archiveprefix = {arXiv}
}

@misc{repoOpenClaw,
  title = {OpenClaw},
  author = {{openclaw}},
  year = {2026},
  howpublished = {\url{https://github.com/openclaw/openclaw}},
  note = {GitHub repository. Accessed 2026-03-13}
}

@misc{repoPicoClaw,
  title = {PicoClaw},
  author = {{sipeed}},
  year = {2026},
  howpublished = {\url{https://github.com/sipeed/picoclaw}},
  note = {GitHub repository. Accessed 2026-03-13}
}

@misc{repoZeroClaw,
  title = {ZeroClaw},
  author = {{zeroclaw-labs}},
  year = {2026},
  howpublished = {\url{https://github.com/zeroclaw-labs/zeroclaw}},
  note = {GitHub repository. Accessed 2026-03-13}
}

@misc{repoNanobot,
  title = {nanobot},
  author = {{HKUDS}},
  year = {2026},
  howpublished = {\url{https://github.com/HKUDS/nanobot}},
  note = {GitHub repository. Accessed 2026-03-13}
}

@misc{repoTinyClaw,
  title = {TinyClaw},
  author = {{TinyAGI}},
  year = {2026},
  howpublished = {\url{https://github.com/TinyAGI/tinyclaw}},
  note = {GitHub repository. Accessed 2026-03-13}
}

@misc{repoNanoClaw,
  title = {NanoClaw},
  author = {{qwibitai}},
  year = {2026},
  howpublished = {\url{https://github.com/qwibitai/nanoclaw}},
  note = {GitHub repository. Accessed 2026-03-13}
}

@misc{repoMoltis,
  title = {Moltis},
  author = {{moltis-org}},
  year = {2026},
  howpublished = {\url{https://github.com/moltis-org/moltis}},
  note = {GitHub repository. Accessed 2026-03-13}
}

@misc{repoIronClaw,
  title = {IronClaw},
  author = {{nearai}},
  year = {2026},
  howpublished = {\url{https://github.com/nearai/ironclaw}},
  note = {GitHub repository. Accessed 2026-03-13}
}

@misc{repoNullClaw,
  title = {NullClaw},
  author = {{nullclaw}},
  year = {2026},
  howpublished = {\url{https://github.com/nullclaw/nullclaw}},
  note = {GitHub repository. Accessed 2026-03-13}
}

@misc{repoMimiClaw,
  title = {MimiClaw},
  author = {{memovai}},
  year = {2026},
  howpublished = {\url{https://github.com/memovai/mimiclaw}},
  note = {GitHub repository. Accessed 2026-03-13}
}

@misc{repoHermitClaw,
  title = {HermitClaw},
  author = {{brendanhogan}},
  year = {2026},
  howpublished = {\url{https://github.com/brendanhogan/hermitclaw}},
  note = {GitHub repository. Accessed 2026-03-13}
}

@misc{repoAstrBot,
  title = {AstrBot},
  author = {{AstrBotDevs}},
  year = {2026},
  howpublished = {\url{https://github.com/AstrBotDevs/AstrBot}},
  note = {GitHub repository. Accessed 2026-03-13}
}

@misc{repoZeptoClaw,
  title = {ZeptoClaw},
  author = {{qhkm}},
  year = {2026},
  howpublished = {\url{https://github.com/qhkm/zeptoclaw}},
  note = {GitHub repository. Accessed 2026-03-13}
}

@misc{repoBabyClaw,
  title = {BabyClaw},
  author = {{yogesharc}},
  year = {2026},
  howpublished = {\url{https://github.com/yogesharc/babyclaw}},
  note = {GitHub repository. Accessed 2026-03-13}
}

@misc{repoSafeClaw,
  title = {safeclaw},
  author = {{princezuda}},
  year = {2026},
  howpublished = {\url{https://github.com/princezuda/safeclaw}},
  note = {GitHub repository. Accessed 2026-03-13}
}

@misc{repoDroidClaw,
  title = {droidclaw},
  author = {{unitedbyai}},
  year = {2026},
  howpublished = {\url{https://github.com/unitedbyai/droidclaw}},
  note = {GitHub repository. Accessed 2026-03-13}
}

@misc{repoFlowlyAI,
  title = {Flowly AI},
  author = {{Nocetic}},
  year = {2026},
  howpublished = {\url{https://github.com/Nocetic/flowlyai}},
  note = {GitHub repository. Accessed 2026-03-13}
}

@misc{repoShrew,
  title = {shrew},
  author = {{Masmedeam}},
  year = {2026},
  howpublished = {\url{https://github.com/Masmedeam/shrew}},
  note = {GitHub repository. Accessed 2026-03-13}
}

@misc{repoZclaw,
  title = {zclaw},
  author = {{tnm}},
  year = {2026},
  howpublished = {\url{https://github.com/tnm/zclaw}},
  note = {GitHub repository. Accessed 2026-03-13}
}

@misc{repoClawlet,
  title = {Clawlet},
  author = {{Kxrbx}},
  year = {2026},
  howpublished = {\url{https://github.com/Kxrbx/Clawlet}},
  note = {GitHub repository. Accessed 2026-03-13}
}

@misc{repoSubZeroClaw,
  title = {subzeroclaw},
  author = {{jmlago}},
  year = {2026},
  howpublished = {\url{https://github.com/jmlago/subzeroclaw}},
  note = {GitHub repository. Accessed 2026-03-13}
}

@misc{repoAutobot,
  title = {Autobot},
  author = {{crystal-autobot}},
  year = {2026},
  howpublished = {\url{https://github.com/crystal-autobot/autobot}},
  note = {GitHub repository. Accessed 2026-03-13}
}

@misc{repoMoxxy,
  title = {moxxy},
  author = {{moxxy-ai}},
  year = {2026},
  howpublished = {\url{https://github.com/moxxy-ai/moxxy}},
  note = {GitHub repository. Accessed 2026-03-13}
}

@misc{repoMicroclaw,
  title = {Microclaw},
  author = {{microclaw}},
  year = {2026},
  howpublished = {\url{https://github.com/microclaw/microclaw}},
  note = {GitHub repository. Accessed 2026-03-13}
}

@misc{repoTroublemaker,
  title = {troublemaker},
  author = {{tinyfatco}},
  year = {2026},
  howpublished = {\url{https://github.com/tinyfatco/troublemaker}},
  note = {GitHub repository. Accessed 2026-03-13}
}

@misc{repoLettaBot,
  title = {LettaBot},
  author = {{letta-ai}},
  year = {2026},
  howpublished = {\url{https://github.com/letta-ai/lettabot}},
  note = {GitHub repository. Accessed 2026-03-13}
}

@misc{repoPicobot,
  title = {picobot},
  author = {{louisho5}},
  year = {2026},
  howpublished = {\url{https://github.com/louisho5/picobot}},
  note = {GitHub repository. Accessed 2026-03-13}
}

@misc{repoAngelClaw,
  title = {AngelClaw},
  author = {{Abdur-rahmaanJ}},
  year = {2026},
  howpublished = {\url{https://github.com/Abdur-rahmaanJ/angel-claw}},
  note = {GitHub repository. Accessed 2026-03-13}
}

@misc{repoSupaClaw,
  title = {SupaClaw},
  author = {{vincenzodomina}},
  year = {2026},
  howpublished = {\url{https://github.com/vincenzodomina/supaclaw}},
  note = {GitHub repository. Accessed 2026-03-13}
}

@misc{repoPickleBot,
  title = {pickle-bot},
  author = {{czl9707}},
  year = {2026},
  howpublished = {\url{https://github.com/czl9707/pickle-bot}},
  note = {GitHub repository. Accessed 2026-03-13}
}

@misc{repoOpenFang,
  title = {OpenFang},
  author = {{RightNow-AI}},
  year = {2026},
  howpublished = {\url{https://github.com/RightNow-AI/openfang}},
  note = {GitHub repository. Accessed 2026-03-13}
}

\end{document}